\documentstyle[emulateapj,epsf]{article}

\begin{document}

\submitted{{\bf PASJ:} Publ. Astron. Soc. Japan {\bf 54}, No. 1, in press.}

\title{A Multi-band Photometric Study of Tidal Debris 
       \\in A Compact Group of Galaxies: Seyfert's Sextet}
\author{Shingo Nishiura$^{1}$, Yasuhiro Shioya$^{2}$, 
        Takashi Murayama$^{2,3,4}$, Yasunori Sato$^{4,5}$, 
        Tohru Nagao$^{2,4}$, Yoshiaki Taniguchi$^{2,4}$, 
        \& D. B. Sanders$^{6}$}
\affil{$^{1}$Kiso Observatory, Institute of Astronomy, The University 
       of Tokyo, Mitake-mura, Kiso-gun, Nagano 397-0101, Japan}
\affil{$^{2}$Astronomical Institute, Graduate School of Science, 
       Tohoku University, Aramaki, Aoba, Sendai 980-8578, Japan}
\affil{$^{3}$Visiting astronomer of Kiso Observatory, 
       University of Tokyo, Japan}
\affil{$^{4}$Visiting astronomer of Mauna Kea Observatories, 
       University of Hawaii, USA}
\affil{$^{5}$Institute of Astronomy, The University of Tokyo, 
       2-21-1, Osawa, Mitaka, Tokyo 181-0015, Japan}
\affil{$^{6}$Institute for Astronomy, University of Hawaii,
       2680 Woodlawn Drive, Honolulu, HI 96822, USA}

\begin{abstract}

In order to investigate the properties of the prominent tidal 
debris feature extending to the northeast of the compact 
group of galaxies Seyfert's Sextet, we analyzed multi-band 
($U$, $B$, $V$, $VR$, $R$, $I$, $J$, $H$ and $K^{\prime}$) 
photometric imaging data and obtained the following results: 
1) The radial surface brightness distribution of this tidal debris 
in Seyfert's Sextet (TDSS) in each band appears to be well 
approximated by an exponential profile.
2) The observed $B-V$ color of TDSS is similar to 
those of dwarf elliptical galaxies in nearby clusters. 
3) Comparing the spectral energy distribution (SED) of TDSS 
with theoretical photometric evolution models and with the SED 
of the stars in the outer part of HCG 79b, 
we find that its SED is comparable to that of a $\sim$ 10 Gyr-old 
stellar population with solar metallicity, similar
to the stellar population in the outer part of HCG 79b. 
This suggests that TDSS consists of stars that may have been 
liberated from HCG 79b by strong galaxy interactions, 
not a pre-existing dwarf galaxy previously thought.   

\end{abstract}


\section{INTRODUCTION}              

Although dwarf galaxies are the most numerous extragalactic objects  
in the nearby Universe (e.g., Ferguson \& Binggeli 1994; 
Binggeli, Sandage \& Tammann 1988; Mateo 1998), it seems 
unclear how they are related in origin to typical large 
(i.e., $L^{*}$) galaxies. Dwarf galaxies could be 
formed through the same formation mechanism as that of 
large galaxies; e.g., gravitational collapse of protogalactic 
gas clouds (Dekel \& Silk 1986; White \& Frenk 1991; 
Frenk et al. 1996; Kauffmann, Nusser \& Steinmetz 1997).
However, it is known that dwarf elliptical galaxies (dEs) 
apparently belong to a different class from normal large 
ellipticals (Es) in the fundamental plane (e.g., Kormendy 1985), 
suggesting that the formation and/or evolution processes of 
dwarfs may not always be the same as those of larger ellipticals. 

It has been argued from an observational view point that
dwarf galaxies may be formed by galaxy collisions because 
there appears to be morphological evidence for dwarf 
galaxies in the tidal tails of interacting galaxies (Zwicky 1956; 
Schweizer 1978; Duc et al. 2000);
i.e., gas-rich dwarf irregular galaxies, can be made out of 
stellar and gaseous material pulled out into intergalactic space by tidal
forces from the disks of colliding parent galaxies.  This possibility has
been recently reinforced by  a number of pieces of observational evidence
(Schweizer 1982;  Bergvall \& Johansson 1985; Schombert, Wallin \&
Struck-Marcell 1990;  Mirabel, Lutz \& Maza 1991; 
Mirabel, Dottori \&  Lutz 1992; Duc \& Mirabel 1994, 1998; 
Duc et al. 2000; Yoshida, Taniguchi \& Murayama 1994; 
Braine et al. 2000; Weilbacher et al. 2000).  Also, Hunsberger et al.
(1996; 1998) find an excess of dwarf galaxies in compact groups of
galaxies apparently caused by interactions among group members. 
Such formation of tidal dwarf galaxies (TDGs) has also
been demonstrated by numerical simulations of merging/interacting 
galaxies (Barnes \& Hernquist 1992; Elmegreen, Kaufman \& 
Thomasson 1993). Therefore, tidal formation seems to 
potentially be an important formation mechanism for 
dwarf galaxies (Okazaki \& Taniguchi 2000 and references therein).

One famous tidal debris system extends to the northeast of 
Seyfert's Sextet (hereafter SS). SS is one of the most famous, 
as well as densest, compact groups of galaxies (Seyfert 1948a, 1948b;
see for a review, Raba\c{c}a 1996). 
This group is also a Hickson compact group (hereafter HCG) of 
galaxies, HCG 79 (Hickson 1982; 1993). 
Many subsequent studies of SS have mentioned that the galaxies in SS appear
to show morphologically and dynamically peculiar properties  (Sulentic \&
Lorre 1983; Rubin, Hunter, \& Ford 1991;  Bettoni \& Fasano 1993; Mendes de
Oliveira \& Hickson 1994;  Bonfanti et al. 1999; Nishiura et al. 2000a). 
Hickson himself regarded SS as a galaxy quartet (HCG 79a, 79b, 79c, 
and 79d).  A fifth component, HCG 79e, was found to be a
redshift-discordant galaxy that is believed to  have no physical relation
to SS (Hickson 1992).   A sixth object, or more precisely the
north-eastern optical fuzz, is now considered likely to be tidal debris 
associated with the morphologically peculiar galaxy HCG 79b (Rubin et al. 
1991; Williams et al. 1991; Mendes de Oliveira \& Hickson 1994; 
V\'{\i}lchez \& Iglesias-P\'{a}ramo 1998).  

Current X-ray observations are not sensitive enough 
to detect any X-ray emission that may originate from
intragroup gas that might be present in SS (Pildis, Bregman, \& Evrard 1995;
Ponman et al. 1996).  However, Sulentic \&  Lorre (1983) and Nishiura et al.
(2000b) detected a faint optical envelope around SS that is plausibly
composed of stars tidally liberated from the galaxies in SS, plus 
Williams, McMahon \& van Gorkom (1991) found extended  H{\sc i}
emission.  These observations suggest that SS is a physically real
compact group.  

In this paper, we present results of our photometric study of TDSS.
Since this tidal debris system in SS (hereafter TDSS) is morphologically 
similar to other tidal debris, such as Arp 105S and Arp 245N 
(Braine et al. 2000), we will also compare the photometric 
properties of TDSS with these two tidal debris. 
Throughout this paper we adopt a distance to SS of 44 Mpc determined using
the mean  recession velocity of HCG 79a, 79b, 79c, and 79d referenced to
the  galactic standard of rest, $V_{\rm GSR}$ = 4449 km s$^{-1}$ 
(de Vaucouleurs et al. 1991), and a Hubble constant, $H_{0}$ = 
100 km s$^{-1}$ Mpc$^{-1}$.
 
\section{DATA}
\subsection{$F336W$, $F439W$, $F555W$, and $F814W$ images}

We obtained the archival HST/WFPC $F439W$ images of SS (PI: Sulentic, 
J. W.). The total exposure time was 8100 seconds (Wu, Raba\c{c}a \& 
Sulentic 1994; Raba\c{c}a 1997). We used IRAF\footnote{Image Reduction 
and Analysis Facility (IRAF) is distributed by the National Optical 
Astronomy Observatories, which are operated by the Association of 
Universities for Research in Astronomy, Inc., under cooperative agreement 
with the National Science Foundation.} to reject cosmic-rays and 
to combine images.

We also obtained the flux-calibrated archival HST/WFPC2 $F336W$, $F439W$, 
$F555W$, and $F814W$ images of SS (PI: Hunsberger, S. D.). The total 
exposure times were 5200 seconds for $F336W$, 5200 seconds for $F439W$, 
2000 seconds for $F555W$, and 2000 seconds for $F814W$.

We compute a Johnson $B$ magnitude assuming that the filter function of 
$F439W$ is the same as that of Johnson {\it B}. We also do a Johnson 
{\it V} magnitude assuming that the filter function of $F555W$ is the 
same as that of Johnson {\it V}. 
According to  Fig 11 in Holtzman et al (1995), 
Johnson {B} $\sim$ {\it B(F439W)} - 0.1 for Johnson {\it B} 
- Johnson {\it V} of about 1. Johnson {V} $\sim$ {\it V(F555W)} - 0.05 
for Johnson {\it V} - {\it I} of about 1. 
Apparent Johnson {\it B} and Johnson {\it V} magnitudes of TDSS may be 
0.1 and 0.05 mag brighter, respectively. 
But, our results using {\it B-V} and {\it V-I} colors do not 
significantly change.

\subsection{$V$ and $R$ images}

$V$- and $R$-band deep images of SS were obtained with 
the 1K (1024 $\times$ 1024) CCD camera attached to the 105 cm Schmidt 
telescope of the University of Tokyo at KISO Observatory, on 20
April, 1996 ($R$-band) and 21 April, 1996 ($V$-band).  The camera provided a
$\approx 12\farcm 5\times 12\farcm 5$ field  of view. The spatial resolution
was 0\farcs 75 per pixel.  The integration time for each exposure was set
to 900 seconds for $V$-band and 600 seconds for $R$-band. 
Four exposures for the $V$-band and five exposures for 
the $R$-band were taken; thus, the total integration time was 3,600 seconds
in the $V$-band image and 3,000 seconds  in the $R$-band image.
The seeing was $\simeq$4\farcs 6 for $V$-band and $\simeq$5\farcs 2 
for $R$-band during the observations.
Data reduction was performed in a standard way using IRAF. 
Flux calibration for the $R$-band images was made using 
the data of photometric standard stars in the field of 
PG 0942$+$029 (Landolt 1992). The $V$-band observations were 
carried out under non-photometric conditions. 
$V$-band image calibration was performed simply by using 
the measured magnitude of a star in the same frame of SS, 
AC2000-738224 (Urban et al. 1997; Kislyuk et al. 1999).  
The photometric errors were estimated to be $\pm 0.27$ mag for 
the $V$-band and $\pm 0.04$ mag for the $R$-band.

\subsection{$VR$ and $I$ images}

We used $VR$- and $I$-band images taken from our previous 
studies (Murayama et al. 2000; Nishiura et al. 2000b). 
Since the $VR$-band is not a standard photometric band 
(Jewitt, Luu \& Chen 1996), we adopted an AB magnitude scale 
for this bandpass. 

\subsection{$J$ and $H$ image}

Near-infrared $J$- and $H$- images of SS were obtained using 
the 105 cm Schmidt telescope at the KISO Observatory during the period 
between 1 July 2001 and 4 July 2001. The telescope was equipped 
with a large-format near-infrared camera called the Kiso Observatory 
Near-Infrared Camera (KONIC) (Itoh et al. 1995). 
KONIC have 1040 $\times$ 1040 pixels providing a $\approx 18\farcm 
\times 18\farcm$ field of view. The spatial resolution was 1.$\farcm$06 
arcsec per pixel. We use the images further binned into 520 $\times$ 520 
pixels, hence a pixel scale is 2$\farcs$12 per the 2$\times$2 binned pixels.

The integration time for each exposure was set to 180 seconds. 
Twenty five exposures for $J$-band and thirty two exposures for 
$H$-band were taken. Therefore the total integration times were 
4500 seconds for $J$-band and 5760 seconds for $H$-band, respectively. 
The seeing was $\sim$ 2$\farcs$5 during the observation.

Data reduction was performed in a standard way using IRAF. 
Flux calibration was made using the measured magnitudes of stars by 
2MASS in the same frame of SS, 1558526$+$203915, 1558546$+$205049, and 
1559287$+$204805. The photometric errors were estimated to be $\pm 0.0x$ 
mag for $J$-band and to be $\pm 0.0x$ mag for $H$-band, respectively. 

\subsection{$K^{\prime}$ image}

Near-infrared {\it K$^{\prime}$}-band images of SS were 
obtained with the 256 $\times$ 256 Infrared Camera attached 
to the f/13.5 Cassegrain focus of the University of 
Hawaii 0.6 m Planetary Patrol telescope at Mauna Kea 
Observatory, on 9 May 1994. 
The camera provided a $\approx 8\farcm 5 \times 8\farcm 5$ field 
of view. The spatial resolution was 2$\farcm$0 arcsec per pixel. 
The integration time for each exposure was set to 120 seconds. 
Eight exposures were taken; thus, the total integration time 
amounted to 960 seconds. 
The seeing was $\sim$ 2$\farcs$4 during the observation.
Data reduction was performed in a standard way using IRAF. 
Flux calibration was made using the data of the UKIRT bright 
standard stars HD84800, HD105601, and HD136754, translated 
into {\it K$^\prime$}-band values (Wainscoat \& Cowie 1992). 
The photometric errors were estimated to be $\pm 0.08$ mag. 

All the images of SS and TDSS are shown in Figure 1. 

\begin{center}
\epsscale{0.5}
\plotone{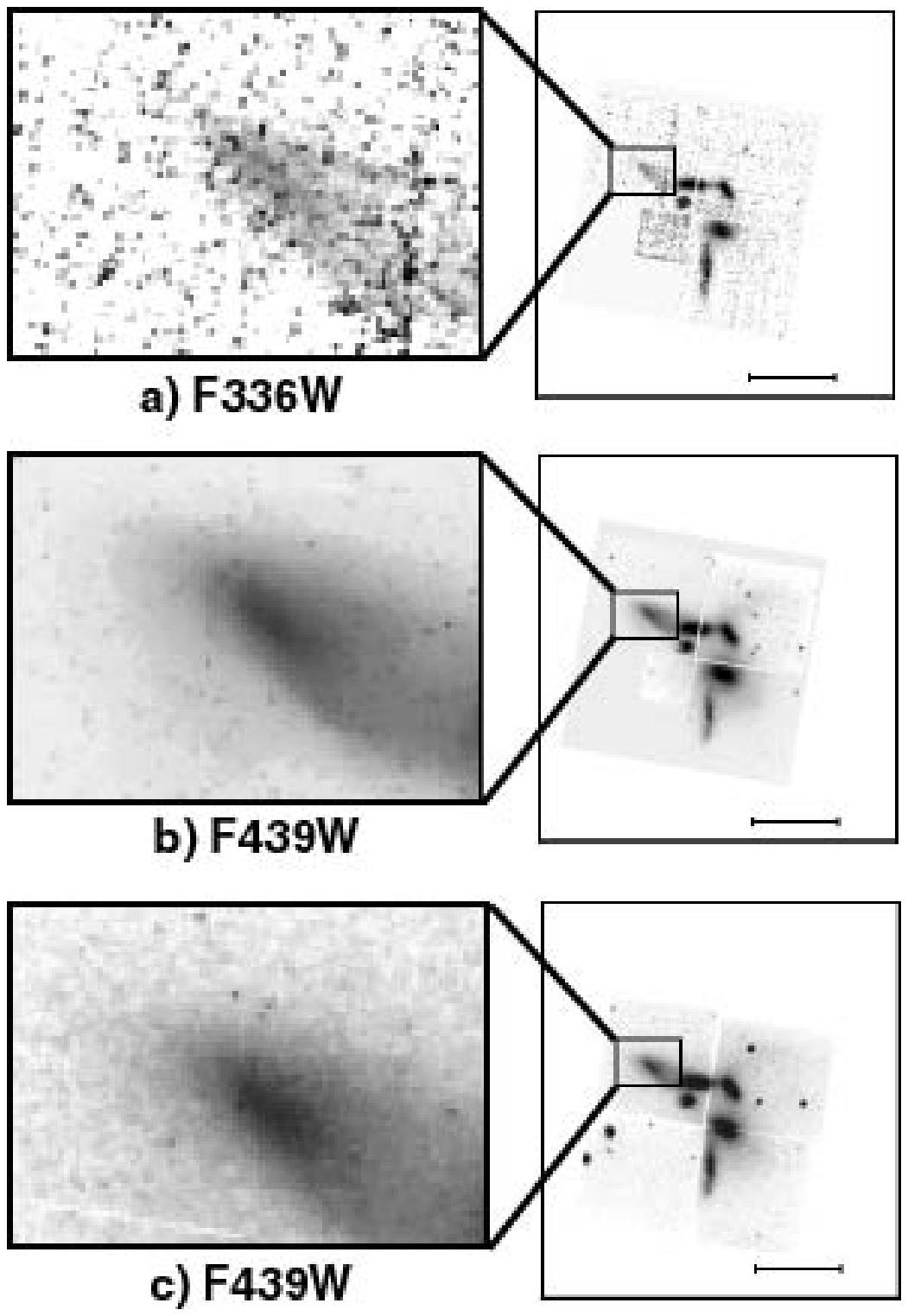}\\
\plotone{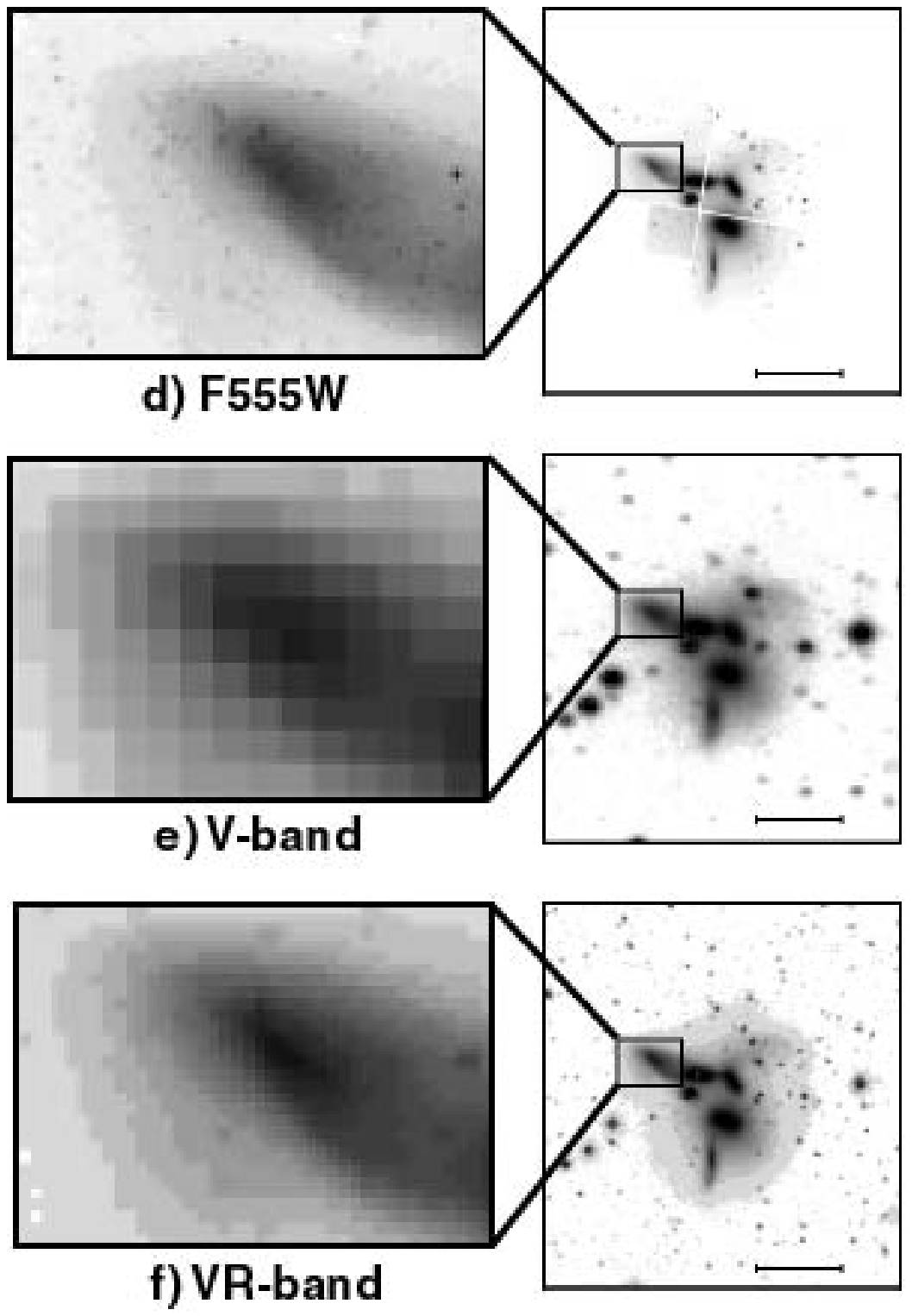}\\
\plotone{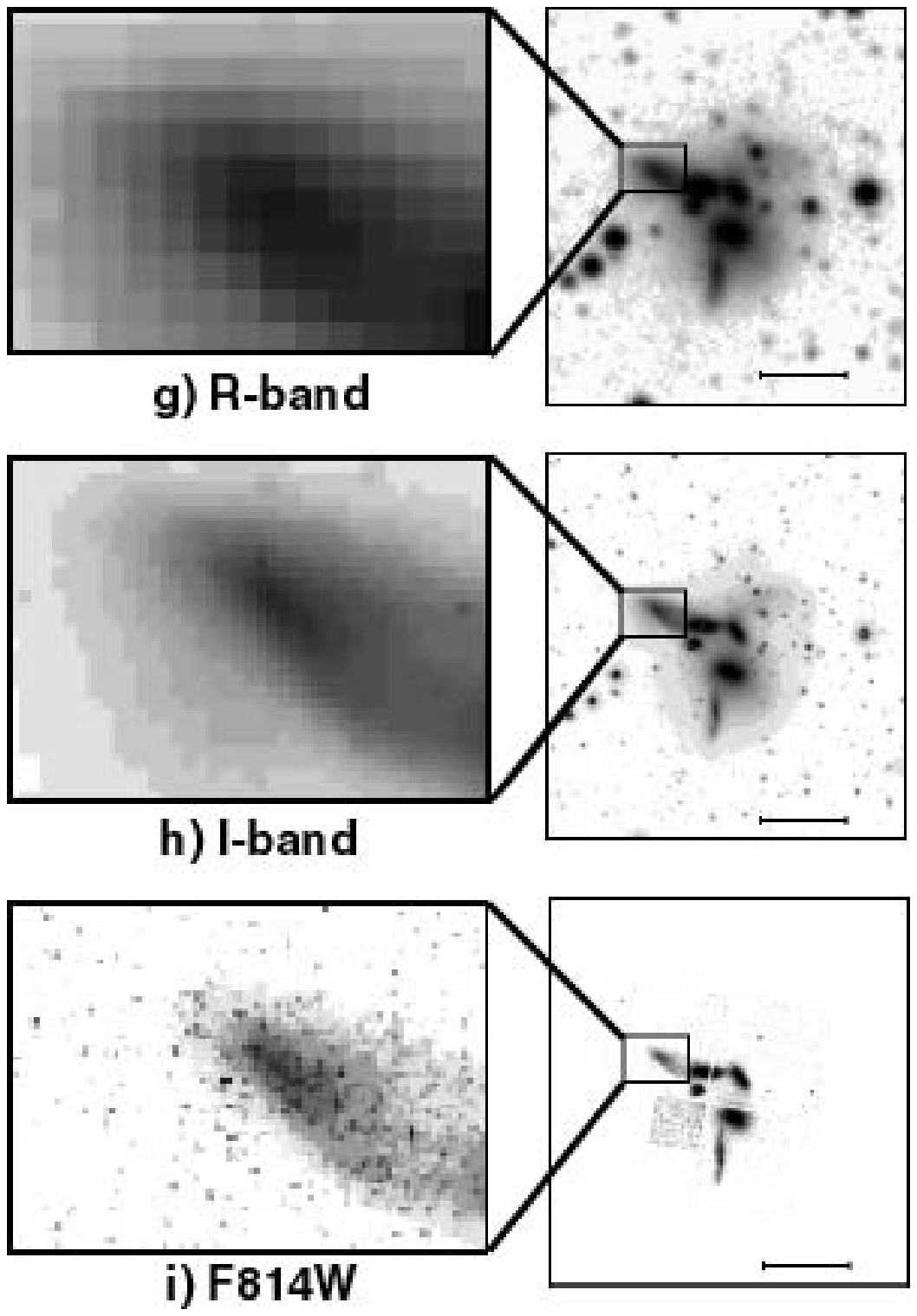}\\
\plotone{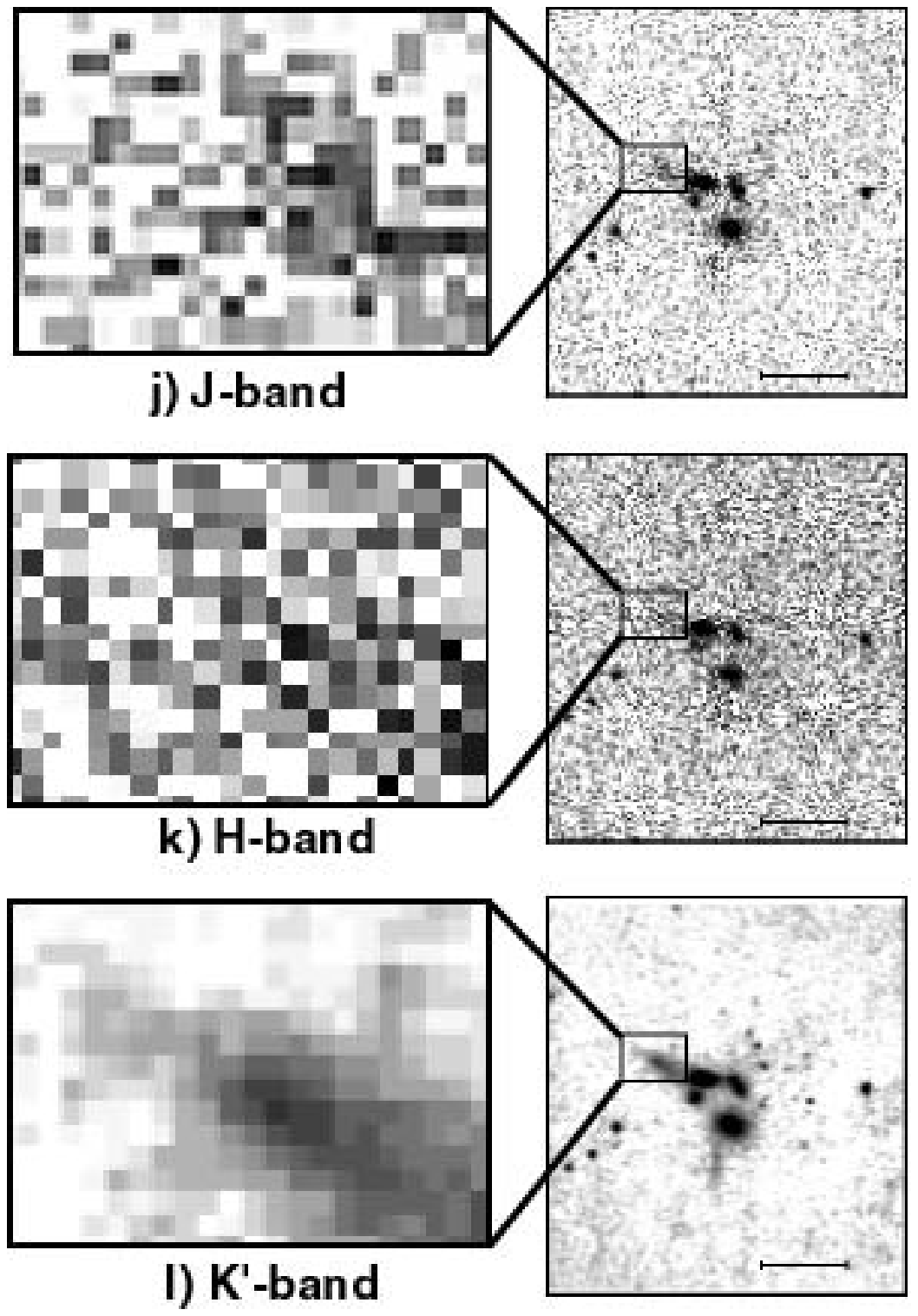}\\
\figcaption{The (a) $F336W$, (b) $F439W$, (c) $F439W/WFPC$, 
(d) $F555W$, (e) {\it V}-band, (f) {\it VR}-band, (g) {\it R}-band, 
(h) {\it I}-band, (i) $F814W$, (j) {\it J}-band, (k) {\it H}-band 
and (l) {\it K$^{\prime}$} images of TDSS and SS before smearing (see text). 
The horizontal bars correspond to 1 arcmin 
$\simeq$ 12.8 kpc. North is up and east is to the left. 
\label{fig1}}
\end{center}

\section{RESULTS}

\subsection{Surface Brightness Profiles of TDSS}

Surface photometry of TDSS was carried out using 
the Surface Photometry Interactive Reduction and Analysis Library 
(SPIRAL: Hamabe \& Ichikawa 1992).
The surface brightness profile along the eastern major axis
in each passband is shown in Figure 2. 
Note that the western part of TDSS may suffer from slight 
contamination by HCG 79b. To minimize light contamination from 
HCG 79b, we show only the eastern-side profile in Figure 2. 
In this procedure, the center of TDSS is defined as the peak 
position in the $VR$- and $I$-band images since our deepest exposures 
are at $VR$- and $I$-band. The peak positions of the two images 
are the same:  $\alpha = 15^{\rm h} 57^{\rm m} 03\fs 2$, 
$\delta = +20\arcdeg 54\arcmin 26\farcs 4$ (B1950.0). 

Unfortunately, the $F336W$, $F439W$ with {\it HST/WFPC2}, $J$- and 
$H$-band surface brightness profiles of TDSS are very low 
signal-to-noise ratio. 
For surface brightness profiles in the other filter bands,  
there appears to be a hump at $r \approx$ 9$^{\prime\prime}$ 
in the {\it F555W}, $VR-$, $I-$ band, and {\it F814W} profiles 
(Figure 2). A broader hump is also seen in the $K^\prime$-band profile.
Since we cannot find  any foreground or background object at 
$r \approx$ 9$^{\prime\prime}$, these humps are considered to be real.
On the other hand, such a hump cannot be seen in the {\it V}- and 
{\it R}-band images. However, since these images were obtained 
in poor seeing conditions, the hump may simply be smeared out. 
The $F439W$ image with {\it HST/WFPC} also appears to be too shallow to 
detect the hump at $r \approx$ 9$\farcs$.  Sulentic \& Lorre (1983), 
Rubin et al. (1991), and Nishiura et al. (2000b) have 
reported that there is a faint optical envelope around SS presumably 
formed by strong galaxy collisions. Therefore, this hump might  
represent the overlap between the faint end of TDSS and
the bright end of the faint optical envelope.

\begin{center}
\epsscale{0.5}
\plotone{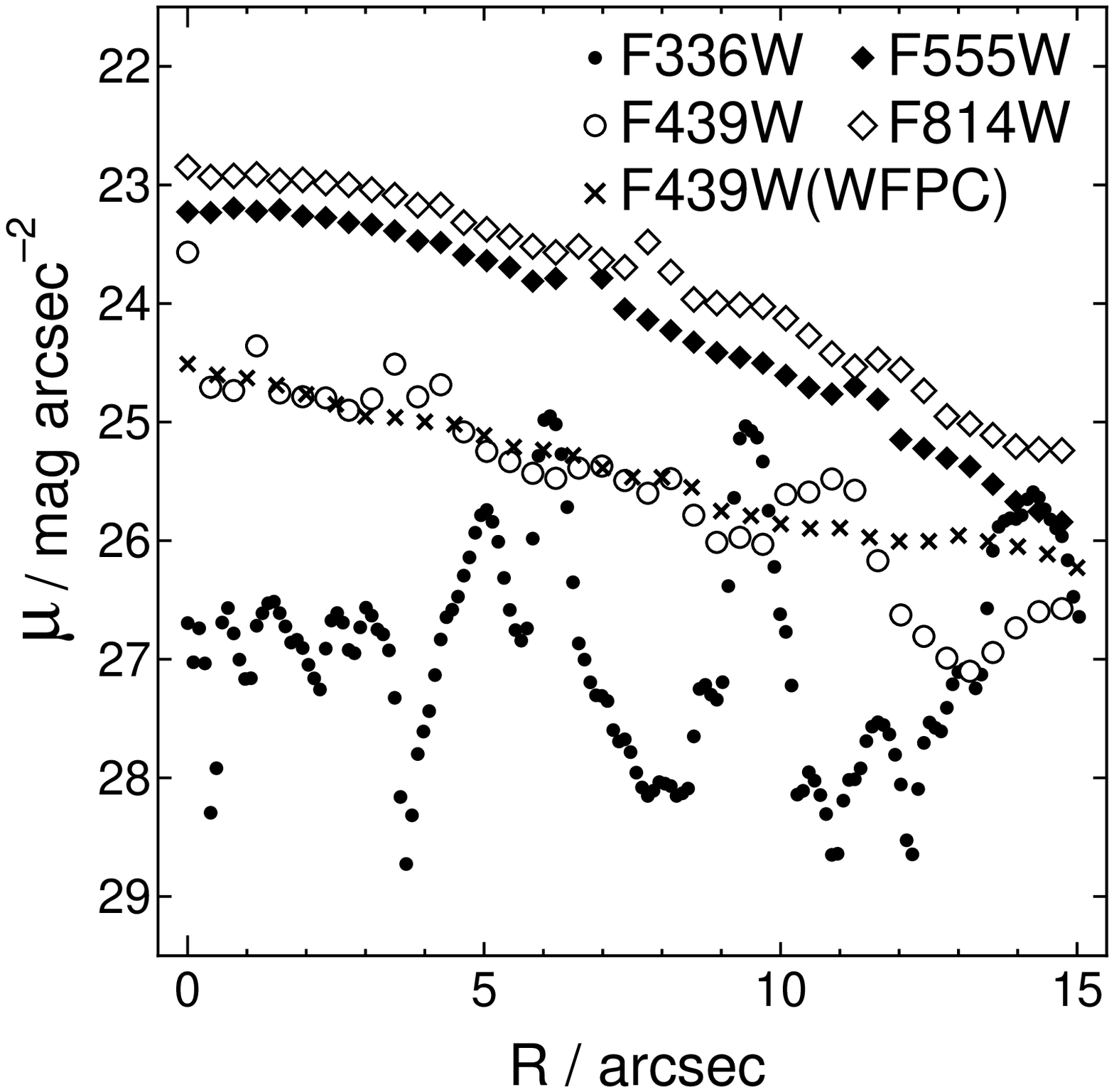}\\
\plotone{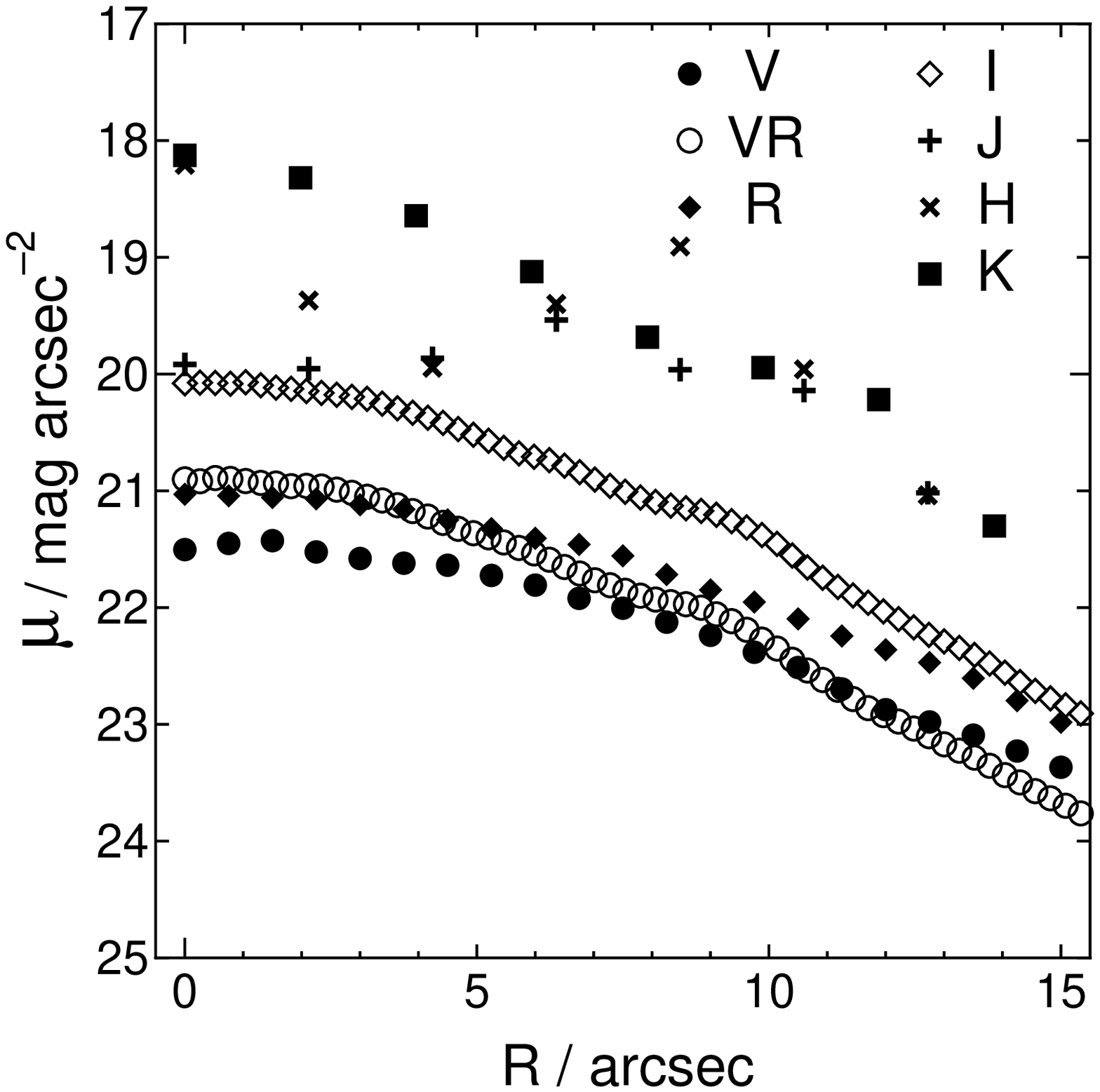}
\figcaption{
The surface brightness (SB) profiles along the major axis 
of TDSS as functions of distance from the center. (a) Filled circles, 
filled diamonds, and open diamonds indicate SB profile in $F336W$, 
$F555W$, and $F814W$, respectively. Open circles mean SB profiles 
in $F439W$ with WFPC2. Crosses represent SB profiles in $F439W$ with WFPC. 
(b) Filled circles, open circles, filled diamonds, and open diamonds 
mean SB profiles in $V$-, $VR$-, $R$-, $I$-band. Pluses, crosses, and 
filled squares indicate SB profiles in $J$-, $H$-, and $K$-band.  
All of SB profiles are without smearing.
\label{fig2}}
\end{center}

The radial surface brightness profiles in the high-quality $VR-$ and 
$I$-band images appear to be well represented by an exponential-like
profile. Such profiles are often observed in dwarf elliptical galaxies 
(Faber \& Lin 1983; Ichikawa et al. 1986a; 1986b), and also in the  
tidal dwarf galaxy Arp 245N (Duc et al. 2000). 

\subsection{Photometric Properties of TDSS}

In order to investigate the stellar content of TDSS,
we now investigate its photometric properties.
We first smear the $F336W$-, $F439W$-, $F555W$-, $V$-, $VR$-, 
$I$-, $F814W$-, $K^{\prime}$-band images to match the seeing of 
our $R$-band image, given that the $R$-band image has the largest 
seeing value in our image set.  
Since a point spread function (PSF) of KONIC is undesirably extended 
(Yanagisawa et al. 1996), for the $J-$ and $H-$ images 
after the PSF deconvolution were carried out with the Lucy-Richardsons 
method realize in the STSDAS packages, we smear both images to match 
the seeing of our $R$-band image.
Then, using the GAIA/SKYCAT\footnote{SKYCAT is a tool that combines 
visualization of images and access to catalogs and archive data for 
astronomy provided by ESO. GAIA is an image display and analysis tool 
that adds many photometry related features to SKYCAT.} package,  
we integrated the light within an ellipse which encloses TDSS in each band;
the semi-major axis = 10$\farcs$5, the eccentricity = 0.90, 
and the position angle = 46$\arcdeg$.
The results are shown in Table 1. 
Unfortunately, in {\it F336W} image large noise overlapped with TDSS, 
we obtained only upper limit of {\it F336W} flux of TDSS. 

We compare the color-magnitude relation found for TDSS with those of 
other similar objects by using a $B-V$ versus $M_{\rm B}$ diagram.  
We obtain an averaged apparent magnitude of 17.13 $\pm$ 0.01 mag 
(Johnson {\it B}) for TDSS after correcting for galactic extinction 
(de Vaucouleurs et al. 1991; Cardelli et al. 1989; Schlegel et al. 1998). 
We also obtained apparent Johnson {\it V} magnitudes of 16.17 $\pm$ 0.27 
(observed with 1K-CCD camera) and of 16.40 $\pm$ 0.003 mag (observed with 
{\it WFPC2}).  Thus, we obtained a Johnson $B-V$ colors for TDSS of 
0.73$\pm$0.01 mag and 0.97$\pm$0.27 mag and an absolute Johnson {\it B} 
magnitude of $-$16.14 mag.

\begin{center}
\epsscale{0.5}
\plotone{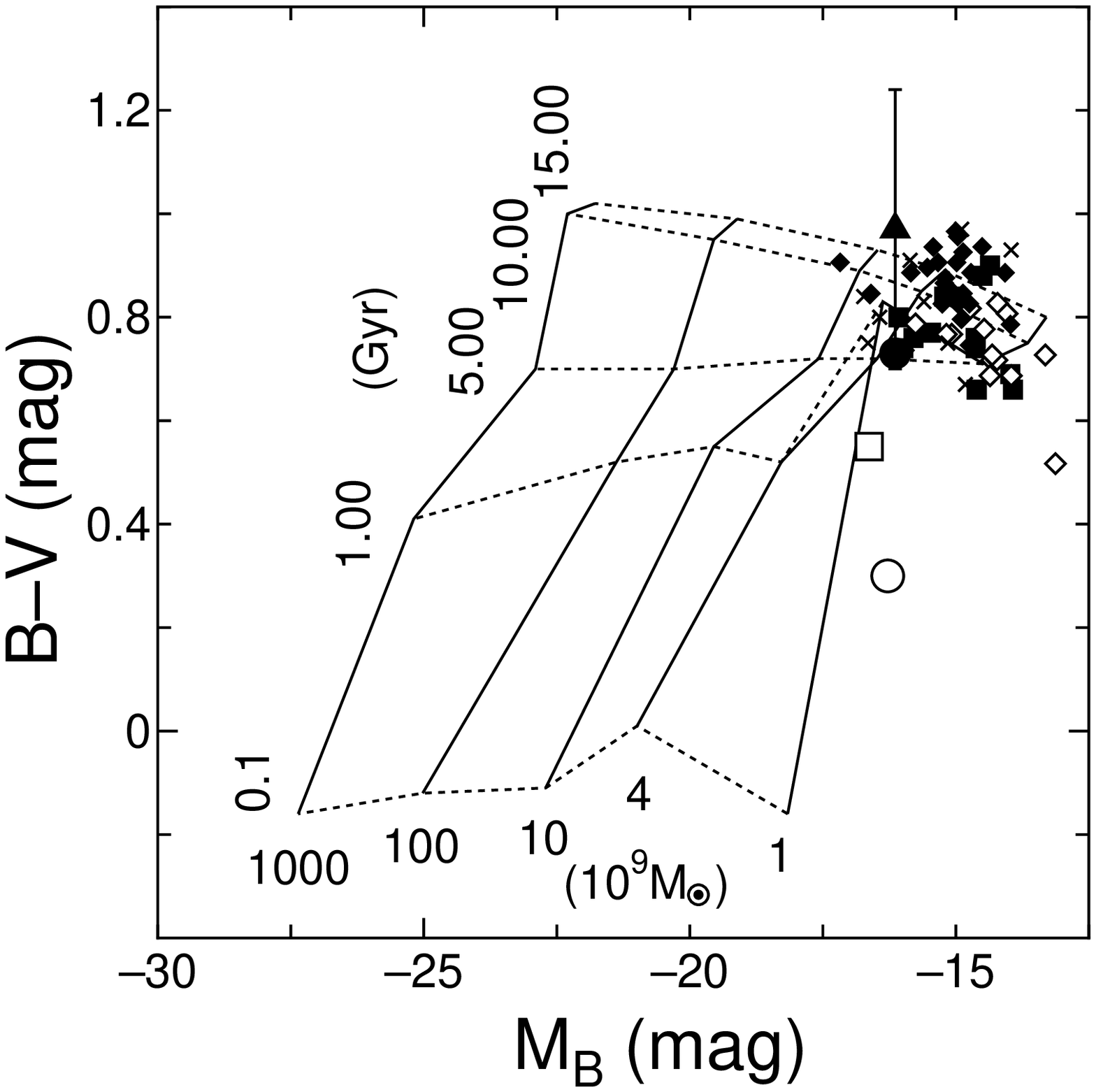}
\figcaption{
Absolute {\it B}-magnitude versus $B-V$ color diagram. 
Filled circle and filled triangle indicate TDSS. 
Open circle and open square indicate 
the tidal dwarf galaxies Arp 105S and Arp 245N, respectively.
Filled squares, open diamonds, filled diamonds, and crosses represent  
dwarf elliptical galaxies in the Virgo cluster, Fornax cluster, Centaurus 
cluster, and NGC 5044 group, respectively. Solid lines and dotted lines 
represent the locus of chemo-photometric evolution models taken from Arimoto
\&  Yoshii (1987): the galaxy mass of $M_{\rm G}/10^{9}M_{\odot}=$1, 4, 10, 
100, and 1000, and the age $t=$0.1, 1, 5, 10, and 15 Gyr. 
\label{fig3}
}
\end{center}

In Figure 3, we plot the data points for TDSS together with
those of two tidal dwarf galaxies, 
Arp 105S ($V_{\rm GSR}=8518$ km s$^{-1}$: de Vaucouleurs et al. 1991) 
and Arp 245N ($V_{\rm GSR}=2175$ km s$^{-1}$, which is the average value
of the $V_{\rm GSR}$ of NGC 2992 and NGC 2993: de Vaucouleurs et al. 1991;
Braine et al. 2000), and dwarf ellipticals in the Virgo cluster (Bothun et
al. 1989),  the Fornax cluster (Bothun et al. 1989), the Centaurus cluster 
(Bothun et al. 1989), and the NGC 5044 group (Cellone 1999).
We also show theoretical loci of chemo-photometric evolution 
models taken from Arimoto \& Yoshii (1987).
The $B-V$ and $M_{B}$ properties of TDSS are different from those of 
Arp 105S ($M_{B}=-16.28$, $B-V=0.3$) and Arp 245N ($M_{B}=-16.63$, 
$B-V=0.55$) (Braine et al. 2000). In Figure 3 TDSS is located near the
bright end of the loci of dwarf ellipticals in nearby clusters. 
Braine et al. (2000) suggested that Arp 105S and Arp 245N 
have experienced recent ($<$ 1 Gyr) bursts of star formation. 
However, the data shown in Figure 3 suggests that a recent ($<$ 1 Gyr) 
star formation burst has not occurred in TDSS. 

\begin{deluxetable}{cccccc}
\tablecaption{Photometric properties of TDSS}
\tablehead{
\colhead{}          & \colhead{$\lambda_{\rm c}$\tablenotemark{a}} 
                    & \colhead{$m_{\rm obs}$\tablenotemark{b}} 
                    & \colhead{$A_{\lambda}$\tablenotemark{c}}     
                    & \colhead{$m_{\rm c}$\tablenotemark{d}} 
                    & \colhead{$f_{\lambda}$} \nl
\colhead{Passband}  & \colhead{(\AA)}   & \colhead{(mag)}         
                    & \colhead{(mag)}   & \colhead{(mag)}    
                    & \colhead{(ergs s$^{-1}$ cm$^{-2}$ ${\rm \AA}^{-1}$)} \nl
}

\startdata
{\it F336W}\tablenotemark{e} & 3317 $\pm$ 185 & $>$17.85 $\pm$ 0.01
                    & 0.30                               & $>$17.56 $\pm$ 0.01
                    & $<$3.45$^{+0.04}_{-0.04} \times 10^{-16}$ \nl
{\it F439W}\tablenotemark{e} & 4283 $\pm$ 232 & 16.80 $\pm$ 0.01 
                    & 0.24                               & 16.56 $\pm$ 0.01 
                    & 8.62$^{+0.05}_{-0.05} \times 10^{-16}$ \nl
                    & 4283 $\pm$ 232    & 16.77 $\pm$ 0.01
                    & 0.24                               & 16.53 $\pm$ 0.01
                    & 8.84$^{+0.06}_{-0.06} \times 10^{-16}$ \nl
{\it F555W}\tablenotemark{e} & 5202 $\pm$ 611 & 16.59 $\pm$ 0.003 
                    & 0.18                               & 16.41 $\pm$ 0.003 
                    & 9.94$^{+0.03}_{-0.03} \times 10^{-16}$ \nl
{\it V}             & 5505 $\pm$ 414    & 16.36 $\pm$ 0.27 
                    & 0.18                               & 16.17 $\pm$ 0.27
                    & 1.22$^{+0.34}_{-0.27} \times 10^{-15}$ \nl
{\it VR$_{\rm AB}$} & 5994 $\pm$ 998    & 16.05 $\pm$ 0.05 
                    & 0.17\tablenotemark{f}              & 15.89 $\pm$ 0.05 
                    & 1.34$^{+0.06}_{-0.06} \times 10^{-15}$ \nl
{\it R$_{\rm c}$}   & 6588 $\pm$ 784    & 15.90 $\pm$ 0.04 
                    & 0.15                               & 15.76 $\pm$ 0.04
                    & 1.07$^{+0.04}_{-0.04} \times 10^{-15}$  \nl  
{\it I$_{\rm c}$}   & 8060 $\pm$ 771    & 15.24 $\pm$ 0.03 
                    & 0.11                               & 15.13 $\pm$ 0.03 
                    & 9.86$^{+0.29}_{-0.28} \times 10^{-16}$ \nl
{\it F814W}\tablenotemark{e} & 8203 $\pm$ 879    & 16.67 $\pm$ 0.003 
                    & 0.11                               & 16.56 $\pm$ 0.003 
                    & 8.63$^{+0.02}_{-0.02} \times 10^{-16}$ \nl
{\it J}             & 12500 $\pm$ 1055 & 14.17 $\pm$ 0.09     
                    & 0.05                               & 14.12 $\pm$ 0.09
                    & 7.55$^{+0.65}_{-0.60} \times 10^{-16}$ \nl     
{\it H}             & 16500 $\pm$ 1490 & 13.52 $\pm$ 0.10    
                    & 0.03                               & 13.48 $\pm$ 0.10 
                    & 4.65$^{+0.45}_{-0.41} \times 10^{-16}$ \nl    
{\it K}\tablenotemark{g} & 22200 $\pm$ 1950 & 13.13 $\pm$ 0.08 
                    & 0.02              & 13.05 $\pm$ 0.08 
                    & 2.35$^{+0.18}_{-0.17} \times 10^{-16}$ \nl
\enddata
\tablenotetext{\rm a}{Errors indicate half value  of the full width 
                      at half maximum for each filter band.}
\tablenotetext{\rm b}{Not corrected for Galactic extinction.}
\tablenotetext{\rm c}{Taken from NASA Extragalactic Database (NED). 
Assuming $A_{B}$=0.237 mag (Schlegel et al. 1998), $E(B-V)$=0.055 
mag, and an $R_{\rm V}$=3.1 extinction curve (Cardelli et al. 1989).}
\tablenotetext{\rm d}{Corrected for Galactic extinction.}
\tablenotetext{\rm e}{Adopted STScI magnitude system (Holtzman et al. 1995).}
\tablenotetext{\rm f}{Adopted simple average of $A_{V}$=0.182 and $A_{R}$ 
=0.147.}
\tablenotetext{\rm g}{Translated from $K^{\prime}$ magnitude (Wainscoat \&
Cowie 1992)}
\end{deluxetable}

Schombert et al. (1990) showed averaged the {\it B-V} color of 
tidal features (tails, bridges, plumes, and envelopes) of 
0.64$\pm$0.24. The {\it B-V} color of TDSS is similar to 
or redder than those of these tidal features. 

\begin{center}
\epsscale{0.5}
\plotone{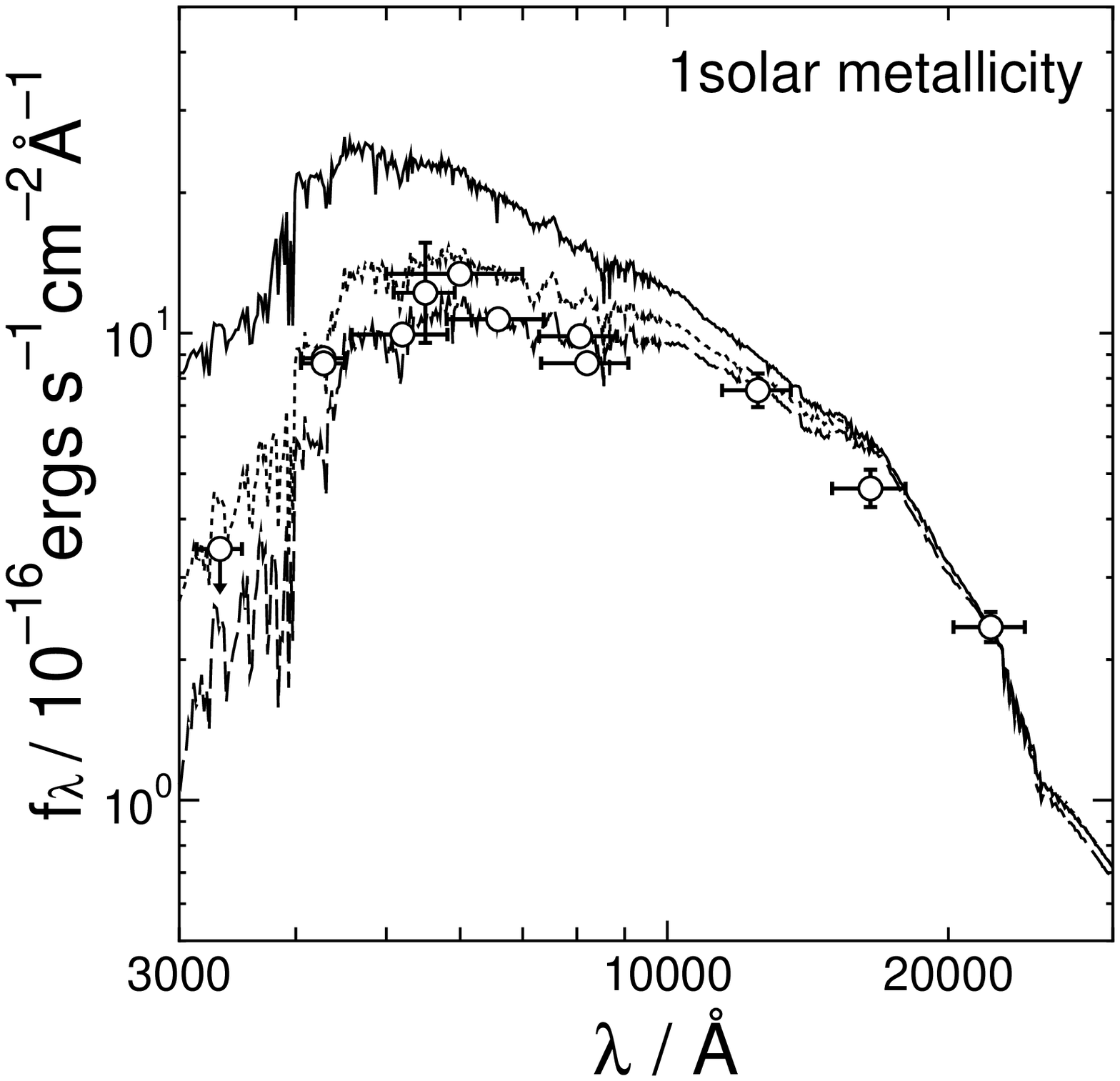}\\
\plotone{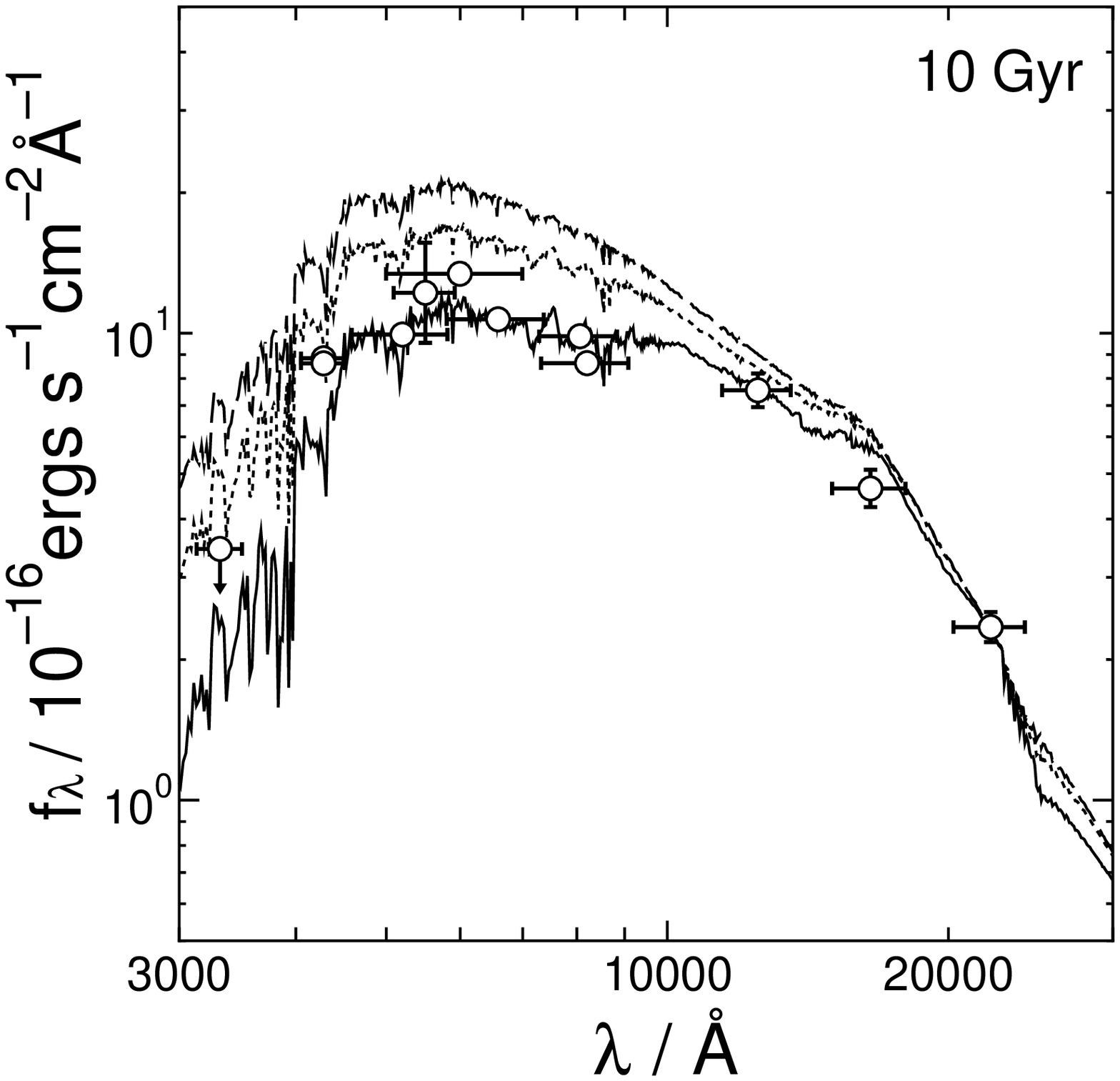}\\
\figcaption{
The spectral energy distribution of TDSS and simple stellar
population models. Filled circles indicate the data for TDSS. Solid,  
dotted, and dashed lines indicate the simple stellar population model - (a)
with solar metallicity at 1, 3, and 10 Gyr,  and (b) the same model at 10
Gyr with 0.2 (dashed line), 0.4 (dotted line),  and 1.0 (solid line) solar
metallicity,  respectively.
\label{fig4}
}
\end{center}

Next, we investigate the spectral energy distribution (SED) of TDSS. 
The observed SED is shown in Figure 4a.
We also show model SEDs generated by GISSEL96 (Bruzual \& Charlot 1993; 
Charlot et al. 1996). Here we adopt the simple stellar population model
with a Salpeter (1955) initial mass function (IMF), solar metallicity, and
upper and lower mass cutoffs of 125 $M_{\odot}$ and 0.1 $M_{\odot}$,
respectively. 
In Figure 4a, we show model SEDs with ages of 1 Gyr, 3 Gyr, and 
10 Gyr. All the models have been scaled to the observed $K^\prime$ flux 
of TDSS for relative comparison to the observed SED.
Furthermore we also show the results of a comparison of the 10 Gyr 
model SEDs with 0.2, 0.4, 1.0 solar metallicity to the observed SED 
in Figure 4b. 
These comparisons suggest that the observed SED is consistent with 
model SEDs of nearly solar metallicity with ages between 3 Gyr and 10 Gyr. 

\section{DISCUSSION}

The main result from our multi-band photometric studies of 
the tidal debris in SS is that TDSS appears to consist
of an old stellar population with $> 0.4$ solar
metallicity (Figure 4a, 4b).  Duc \& Mirabel (1999) indicate that tidal
dwarf galaxies have  an average metallicity of one third of solar,
which is independent of their absolute blue magnitude 
(Duc et al. 2000). The metallicity of TDSS thus appears to be higher 
with compared to tidal dwarf galaxies. 

\begin{center}
\epsscale{0.5}
\plotone{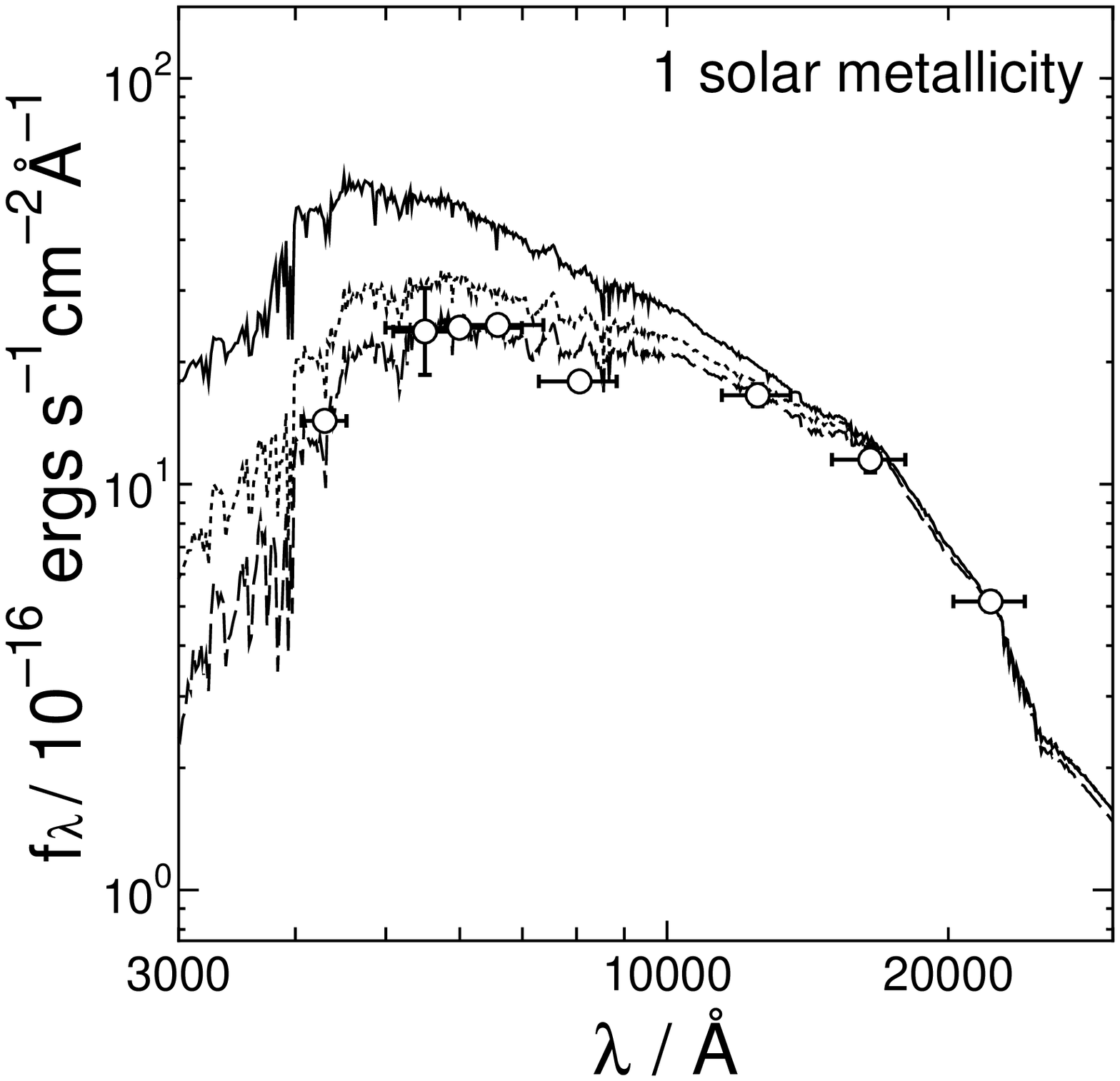}
\figcaption{
The spectral energy distribution of the outer part of 
HCG 79b and simple stellar
population models. Filled circles indicate the data of HCG 79b. Solid,  
dotted, and dashed lines indicate the simple stellar population model 
with a solar metallicity at $t=$1, 3, and 10 Gyr, respectively. 
\label{fig5}
}
\end{center}

\begin{deluxetable}{cccccc}
\tablecaption{Photometric properties of the outer part of HCG 79b}
\tablehead{
\colhead{}  & \colhead{$\lambda_{\rm c}$\tablenotemark{a}} 
                    & \colhead{$m_{\rm obs}$\tablenotemark{b}} 
                    & \colhead{$A_{\lambda}$\tablenotemark{c}}     
                    & \colhead{$m_{\rm c}$\tablenotemark{d}} 
                    & \colhead{$f_{\lambda}$} \nl
Passband            & \colhead{(\AA)}    & \colhead{(mag)}         
                    & \colhead{(mag)}    & \colhead{(mag)}    
                    & \colhead{(ergs s$^{-1}$ cm$^{-2}$ ${\rm \AA}^{-1}$)} \nl
}
\startdata 
{\it F439W}\tablenotemark{e} & 4283 $\pm$ 232 & 16.28 $\pm$ 0.01 
                    & 0.24                               & 16.04 $\pm$ 0.01 
                    & 1.39$^{+0.01}_{-0.01} \times 10^{-15}$ \\
{\it V}             & 5505 $\pm$ 414    & 15.63 $\pm$ 0.27 
                    & 0.18                               & 15.45 $\pm$ 0.27
                    & 2.38$^{+0.66}_{-0.52} \times 10^{-15}$ \\
{\it VR$_{\rm AB}$} & 5994 $\pm$ 998    & 15.41 $\pm$ 0.05 
                    & 0.17\tablenotemark{f}              & 15.24 $\pm$ 0.05 
                    & 2.43$^{+0.12}_{-0.11} \times 10^{-15}$ \\
{\it R$_{\rm c}$}   & 6588 $\pm$784     & 15.00 $\pm$ 0.04 
                    & 0.15                               & 14.85 $\pm$ 0.04
                    & 2.47$^{+0.08}_{-0.08} \times 10^{-15}$  \\  
{\it I$_{\rm c}$}   & 8060 $\pm$ 771    & 14.58 $\pm$ 0.03 
                    & 0.11                               & 14.48 $\pm$ 0.03 
                    & 1.79$^{+0.05}_{-0.05} \times 10^{-15}$ \\
{\it J}             & 12500 $\pm$ 1055  & 13.31 $\pm$ 0.07    
                    & 0.05                               & 13.26 $\pm$ 0.07
                    & 1.66$^{+0.11}_{-0.10} \times 10^{-15}$ \\
{\it H}             & 16500 $\pm$ 1490  & 12.53 $\pm$ 0.08    
                    & 0.03                               & 12.50 $\pm$ 0.08
                    & 1.15$^{+0.09}_{-0.08} \times 10^{-15}$ \\
{\it K}\tablenotemark{g} & 22200 $\pm$ 1950 & 12.22 $\pm$ 0.06 
                    & 0.02              & 12.20 $\pm$ 0.06 
                    & 5.14$^{+0.30}_{-0.28} \times 10^{-16}$ \\
\enddata
\tablenotetext{\rm a}{Errors indicate half value of the full width 
                      at half maximum for each filter band.}
\tablenotetext{\rm b}{Not corrected for Galactic extinction.}
\tablenotetext{\rm c}{Taken from NASA Extragalactic Database (NED). 
Assuming $A_{B}$=0.237 mag (Schlegel et al. 1998), $E(B-V)$=0.055 
mag, and an $R_{\rm V}$=3.1 extinction curve (Cardelli et al. 1989).}
\tablenotetext{\rm d}{Corrected for Galactic extinction.}
\tablenotetext{\rm e}{Adopted STScI magnitude system (Holtzman et al. 1995).}
\tablenotetext{\rm f}{Adopted simple average of $A_{V}$=0.182 and $A_{R}$ 
=0.147.}
\tablenotetext{\rm g}{Translated from $K^{\prime}$ magnitude (Wainscoat \&
Cowie 1992)}
\end{deluxetable}

The spatial configuration of TDSS and the member galaxies in SS 
and the spatial distribution of H{\sc i} emission suggest 
that TDSS was liberated from HCG 79b during the strong interaction 
of HCG 79b with HCG 79d (Williams et al. 1991; Mendes de Oliveira \& 
Hickson 1994). 
Therefore, it seems reasonable to consider that TDSS mainly consists 
of stars liberated from the outer region of HCG 79b. 
This means that the old stellar population 
with nearly solar metallicity in TDSS should be identical to 
that of the outer region of HCG 79b.
In order to investigate the stellar population of the outer region of 
HCG 79b, we carried out aperture photometry of the outer region of
HCG 79b by integrating the light within an annular ellipse 
with semi-major axes between 5$\farcs$2 and  12$\farcs$0. 
The lower cutoff radius is determined to avoid the inner 
star forming regions of HCG 79b (V\'{\i}lchez \& Iglesias-P\'{a}ramo 
1998; Shimada et al. 2000).
The ellipse eccentricity and position angle are set to be 0.85 
and 8$\arcdeg$, respectively.
The center coordinate of the annular ellipse is
$\alpha = 15^{\rm h} 57^{\rm m} 00\fs 8$,
$\delta = +20\arcdeg 54\arcmin 15\farcs 4$ (B1950.0) (Hickson 1993). 
In four {\it WFPC2} images, since lines of insensitive pixels 
run through the HCG79b, we cannot obtain the four fluxes ({\it F336W}, 
{\it F439W}, {\it F555W}, {\it F814W}) of TDSS with {\it WFPC2}. 
The resultant SED is shown in Figure 5 (see also Table 2). 
The same model SEDs used in Figure 4a 
are also shown in Figure 5.  Although the $I$-band flux is a bit
weaker than the model one, the overall SED is well described 
by a model SED with an age of 10 Gyr and solar metallicity. 
>From these SED comparisons, it does seem likely that TDSS was liberated
from  HCG 79b by strong interaction with HCG 79d. This would then be the
reason why the abundance of TDSS is richer that those of tidal dwarf 
galaxies. Schombert et al. (1990) has also shown that the colors of many 
tidal features are similar to the outer parts of the parent galaxies 
further strengthening the ``stripping origin" hypothesis. 

Comparing the spatial distribution 
of H{\sc i} emission with the numerical
simulation by Toomre \& Toomre (1972), Williams et al (1991) suggested
that the H{\sc i} countertail and the optical counterarm from HCG 79b
could have been generated $\approx$ 0.5 Gyr ago after the  
close encounter between HCG 79b and HCG 79d.
Since it is known that tidal arms are promptly liberated from
interacting galaxies, it seems likely that a secondary star 
formation event could have been induced during the early phase of 
tidal liberation such as appears to be the case in tidal dwarf
galaxies (Figure 3). 

\begin{center}
\epsscale{0.5}
\plotone{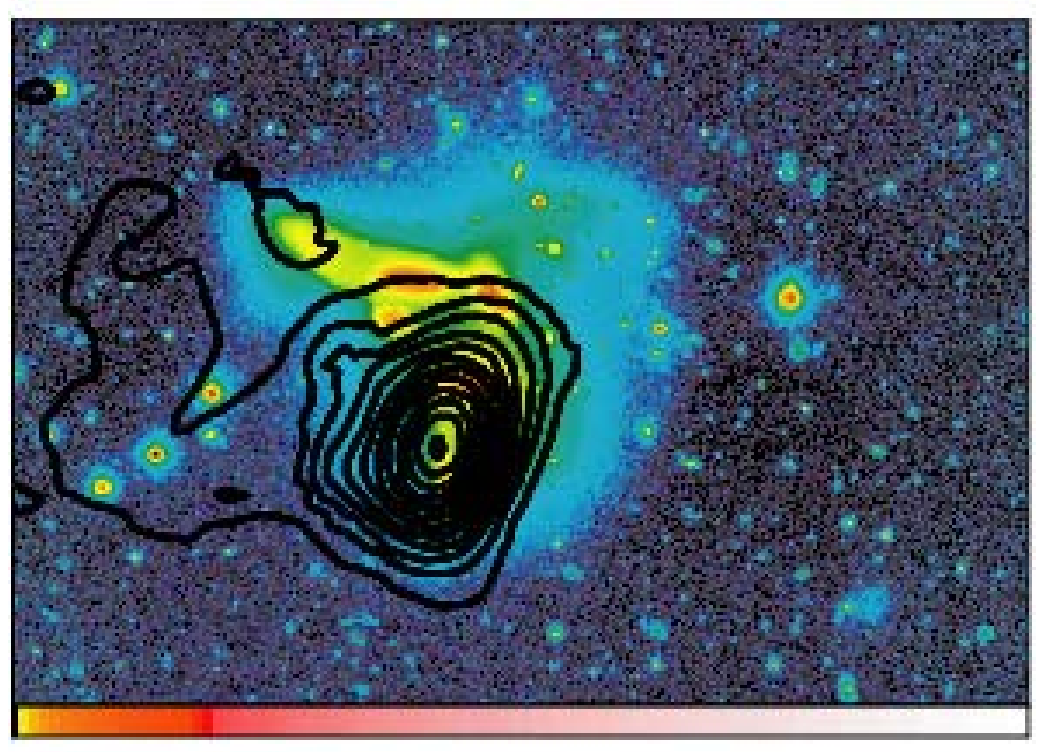}
\figcaption{
Optical $VR+I$-band image and the intensity map of 
the H{\sc i} emission taken from Williams et al. (1991).  H{\sc i} 
contours are drawn at levels 0.05, 0.15, 0.25, 0.35, 0.45, 0.55, 0.65, 
0.75, 0.85, 0.95, 1.05, and 1.15 Jy km s$^{-1}$ per beam. 
\label{fig6}
}
\end{center}

Williams et al. (1991) detected H{\sc i} emission 
at a level of $1.4 \times 10^{20}$ 
H cm$^{-2}$ at the eastern end of TDSS. 
In Figure 6, we show our optical $VR+I$ image together with 
the H{\sc i} 
intensity map taken from Williams et al. (1991).
The H{\sc i} emission appears associated with TDSS and distributed 
in an ellipse 
whose semi-major and semi-minor axes are 15$\farcs$0  and 10$\farcs$0, 
respectively.
If we assume that the H{\sc i} column density within 
this ellipse is constant at a value of $1.4 \times 10^{20}$ 
H cm$^{-2}$,   
we roughly estimate that the H{\sc i} gas mass associated 
with TDSS is $2 \times 10^{7} M_{\odot}$. Thus, we obtain a ratio 
of the H{\sc i} gas mass to optical $B-$band luminosity 
of $3.2 \times 10^{-2} M_{\odot}/L_{\odot}$ for TDSS. 
This value is much smaller than those of 
Arp 105S ($L_{B}=4.9\times 10^{8}L_{\odot}$, $M_{\rm H{\sc i}}
=4\times 10^{8}M_{\odot}$, $M_{\rm H{\sc i}}/L_{B}=0.8
M_{\odot}/L_{\odot}$) and Arp 245N ($L_{B}=6.1\times 10^{8}
L_{\odot}$, $M_{\rm H{\sc i}}=6\times 10^{8}M_{\odot}$, $M_{\rm H{\sc i}}
/L_{B}=1.0M_{\odot}/L_{\odot}$). 
Furthermore, V\'{\i}lchez \& Iglesias-P\'{a}ramo (1998) detect 
no H$\alpha$ emission in TDSS. 
This suggests the poverty of very massive stars in TDSS. 
These observations suggest that only a small amount of gas was 
involved in the formation of TDSS. 

Bettoni \& Fasano (1993) and Bonfanti et al. (1999) mentioned  
another possibility; i.e., TDSS (called ``HCG 79b1'' in their papers)
is a pre-existing galaxy falling into SS that is being destroyed by 
galaxy-galaxy interaction. 
Since the photometric properties of TDSS are similar to those of
dEs (i.e., B-V colors and exponential surface brightness profiles),
their idea may remain as a possible one.  
However, from the metallicity of TDSS, as determined 
from the SED, we suggest that TDSS is a tidally-induced object
consisting primarily of stars liberated from HCG 79b. Further investigation 
of the mass-to-luminosity ratio and/or a more detailed
investigation of the metallicity would be useful to confirm this hypothesis.
It also requires numerical simulations to make sure whether tidally object 
being made by galaxy interactions without secondary star formation.

\section{CONCLUSIONS}

From multi-band photometry of the prominent tidal debris 
feature to the northeast of Seyfert's Sextet we have obtained 
the following results: 
\begin{enumerate}
\item The surface brightness profile of this tidal debris 
      in Seyfert's Sextet (TDSS) in each band shows 
      an approximately exponential profile.
\item The observed $B-V$ color of TDSS is redder than  
      those of tidal dwarf galaxies, e.g., Arp 105S 
      and Arp 245N (Braine et al. 2000). 
\item Comparing the spectral energy distribution (SED) of TDSS 
      with theoretical photometric evolution models, 
      we find that its SED is comparable to that of a stellar 
      population with age $\sim$ 10 Gyr and with higher  
      metallicity than the average values found in tidal dwarf
      galaxies.
\item Comparing SEDs, we find that the SED of TDSS is similar to 
      that of the stars in the outer part of HCG 79b. 
\item TDSS seems to consist primarily of stars liberated from 
      HCG 79b by strong galaxy interaction.
\end{enumerate}

We conclude that TDSS is simply a passive tidal feature like 
many others in interacting galaxy systems. We also conclude 
that there is no indication that TDSS will evolve into 
a star-forming tidal dwarf galaxy similar to those previously 
studied. This, however, indicates that another type of 
forming dwarf galaxy without secondary star formation through 
the galaxy interaction.  

\vspace{1ex} 

We would like to thank the staff members of the Okayama 
Astrophysical Observatory, the KISO observatory and 
the UH 2.2 m telescope for their kind assistance during 
our observations. We thank J. W. Sulentic for his HST observation 
of Seyfert's Sextet, and an anonymous referee for several useful comments
which helped improve the paper. We also thank Ichi Tanaka for his kind help 
during our observations, Richard Wainscoat and Shinki Oyabu  for their
useful comments on photometric calibration,  and Daisuke Kawata for useful
discussions.  A part of this work was done when YT was a visiting
astronomer  at the IfA, University of Hawaii. YT would like to thank
Rolf-Peter  Kudritzki and Bob McLaren for their warm hospitality. 
YS thanks the Japan Society for Promotion of Science (JSPS) 
Research Fellowships for Young Scientist. 
This work was supported in part by the Ministry of Education, Science,
Sports and Culture in Japan under Grant Nos. 07055044, 10044052, 
and 10304013.


\end{document}